\documentclass[article,twocolumn,english,secnumarabic,footinbib,tightenlines,nobibnotes,aps,prl,unsortedaddress,showpacs,superscriptaddress]{revtex4-1}
\usepackage{graphicx}
\usepackage[range-units=single,separate-uncertainty]{siunitx}
\usepackage[color links=true,linkcolor=blue,citecolor=blue,urlcolor=blue]{hyperref}

\usepackage[usenames,dvipsnames]{color}
\usepackage{amsmath}
\usepackage{amsfonts}
\usepackage{amssymb}
\usepackage{mathtools}
\usepackage{lipsum}
\usepackage{pict2e}
\usepackage{hyperref}
\usepackage{graphicx,xcolor}
\usepackage{bbding}
\usepackage{bm}
\usepackage{bbm}

\usepackage{multirow}
\usepackage{dcolumn}  
\usepackage{float}
\usepackage{amsmath,mathrsfs}
\usepackage{amsthm}
\usepackage{epsfig}

\usepackage{array}
\usepackage{color}
\usepackage{soul}
\usepackage[caption=false]{subfig}

\usepackage{comment}

\newcolumntype{P}[1]{>{\centering\arraybackslash}p{#1}}

\mathchardef\sOmega="710A
\mathchardef\sGamma="7100
\mathchardef\sDelta="7101

\def\frac#1#2{{\textstyle{#1 \over #2}}}

\newcommand{\nvec}[1]{\textbf{#1}}

\newcommand{\im}{i}

\newcommand{\psmatrix}[2][1.2]{%
	\scalebox{#1}{%
		\renewcommand{\arraystretch}{1.2}%
		$\begin{pmatrix}#2\end{pmatrix}$%
	}
}

\begin{document}
	
	\title{Band structure engineering and reconstruction in electric circuit networks}
	
	\author{Tobias Helbig}
        \thanks{Both authors equally contributed to the work.}
	\affiliation{Institute for Theoretical Physics and Astrophysics, University of W\"urzburg, D-97074 W\"urzburg, Germany}
	\author{Tobias Hofmann}
        \thanks{Both authors equally contributed to the work.}
	\affiliation{Institute for Theoretical Physics and Astrophysics, University of W\"urzburg, D-97074 W\"urzburg, Germany}
	\author{Ching Hua Lee}
	\affiliation{Department of Physics, National University of Singapore, Singapore, 117542.}
	\affiliation{Institute of High Performance Computing, A*STAR, Singapore, 138632.}
	\author{Ronny Thomale}
	\email{Corresponding author: rthomale@physik.uni-wuerzburg.de}
	\affiliation{Institute for Theoretical Physics and Astrophysics, University of W\"urzburg, D-97074 W\"urzburg, Germany}

	\author{Stefan Imhof}
	\affiliation{Physikalisches Institut der Universit\"at W\"urzburg, 97074 W\"urzburg, Germany}
	\author{Laurens W. Molenkamp}
	\affiliation{Physikalisches Institut der Universit\"at W\"urzburg, 97074 W\"urzburg, Germany}
	\author{Tobias Kiessling}
	\affiliation{Physikalisches Institut der Universit\"at W\"urzburg, 97074 W\"urzburg, Germany}
	
	\date{\today}
	
	\begin{abstract}
		We develop an approach to design, engineer, and measure band structures in a synthetic crystal composed of electric circuit elements. Starting from the nodal analysis of a circuit lattice in terms of currents and voltages, our Laplacian formalism for synthetic matter allows us to investigate arbitrary tight-binding models in terms of wave number resolved Laplacian eigenmodes, yielding an admittance band structure of the circuit. For illustration, we model and measure a honeycomb circuit featuring a Dirac cone admittance bulk dispersion as well as flat band admittance edge modes at its bearded and zigzag terminations. We further employ our circuit band analysis to measure a topological phase transition in the topolectrical Su-Schrieffer-Heeger circuit. 
	\end{abstract}

	
	\maketitle

	{\it Introduction.} Electrons in a periodic lattice potential~\cite{bloch,Kronig499} is one of the most central problems in the history of condensed matter research. As our understanding of it progressed over the decades, revolutionary concepts have kept arising from there such as, most recently, relativistic particle dispersions in graphene~\cite{castroneto-09rmp109} or topologically non-trivial insulators and semimetals~\cite{RevModPhys.82.3045,RevModPhys.83.1057,RevModPhys.90.015001}. In this context, synthetic matter has emerged as a complementary branch to realize lattice potential environments for alternative degrees of freedom. This includes, among others, atoms in optical lattices, exciton-polaritons in semiconductor platforms, photons in cavities and waveguides, mechanical and acoustic settings, and several more~\cite{RevModPhys.80.885,lailai,Knight:96,marin,PhysRevLett.114.114301,trunk}. The common purpose of synthetic matter research is to either accomplish a highly tunable simulator for a given electronic lattice problem, or to establish a framework in which an intricate lattice model can be experimentally realized in the first place.
	
	Electric circuit networks~\cite{kirchhoff,circuits} naturally present themselves as yet another physical system in which a lattice potential along with tunable lattice connectivity can be realized. While most applications in electrical engineering do not specifically necessitate a translationally invariant arrangement of circuit elements, electric circuit networks still represent a prototypical candidate for such synthetic matter. In the realm of topological matter, it has recently been discovered that a two-dimensional topological crystalline insulator can be built in an electric circuit~\cite{PhysRevLett.106.106802,ningyuan2015time,PhysRevLett.114.173902}, which was subsequently generalized to the prescription for modelling topological insulators, topological semimetals, and higher-order topological states of arbitrary dimension in topolectrical circuits~\cite{ustopo,ustopo2}.
	
	In this Letter, we develop the framework to build and measure admittance band structures in an electric circuit in a way that allows for a precise translation from a given tight-binding model to its circuit realization. We employ a Laplacian formalism put forward by us~\cite{ustopo} to connect the node-wise currents of the circuit with the node-wise voltages measured against ground. For a translationally invariant system, the circuit Laplacian, whose eigenvalues form the circuit admittance spectrum, then inherits a block diagonal form due to a wave number component $k$ per periodic direction. As such, the energy band structure from a given abstract tight-binding model translates into an admittance band structure for the circuit derived from the Fourier analysis of site-resolved voltages and currents, lending itself to immediate measurability. The reconstruction of the band structure is thereby for the first time straightforwardly accessable in a systematic and scalable measurement in terms of an electrical circuit environment. We illustrate our admittance band engineering for a two-dimensional periodic circuit lattice reminiscent of graphene, its different surface terminations, and the topological phase transition in the Su-Schrieffer-Heeger (SSH) model as a function of the ratio between the intracell and intercell hopping amplitude.

	\begin{figure*}[t]
		\includegraphics[width=\linewidth]{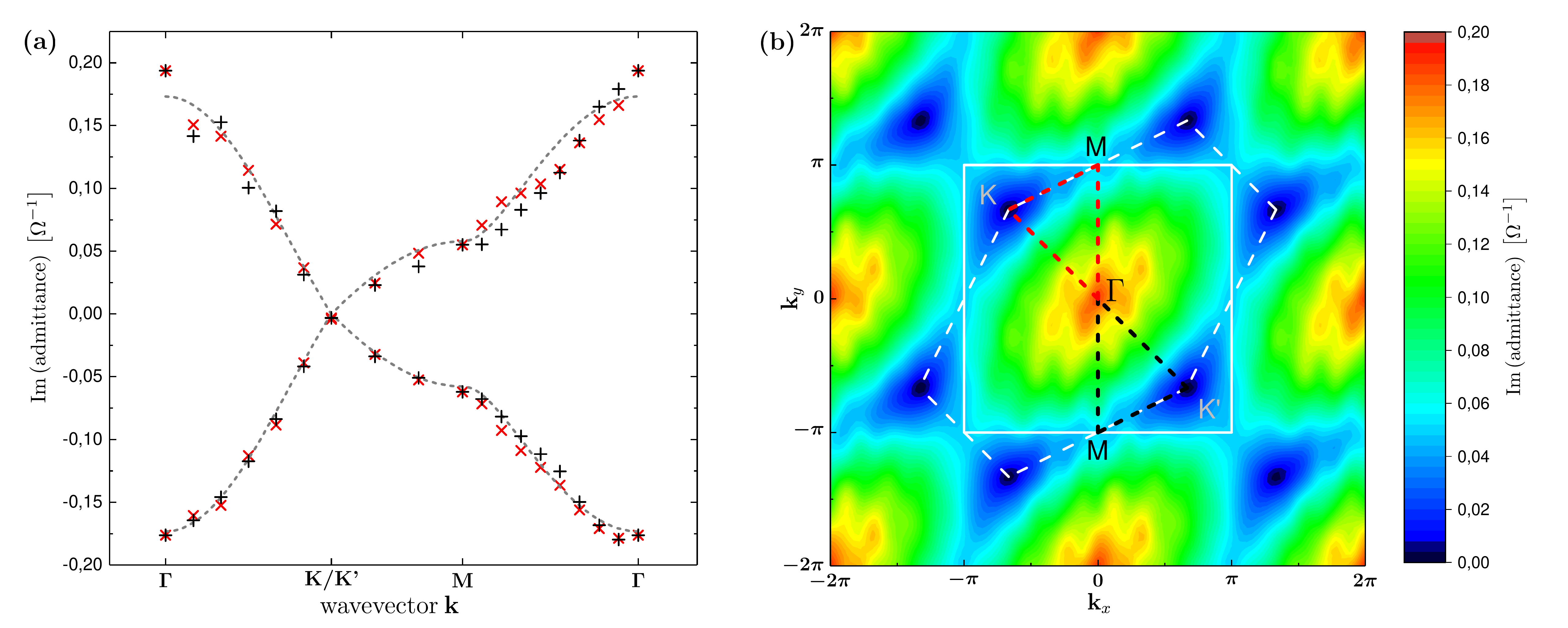}
		\caption{(a) Admittance band structure of the honeycomb circuit (two-node unit cell shown in the upper central inset).  Grey dashed lines highlight the continuum theoretical admittance dispersion. In the absence of any fixed length scale entering the circuit system of connected nodes, there is an equivalence class of individually scaled and oriented (reciprocal) lattice vectors as long as lattice connectivity and number of nodes per unit cell is preserved. The red $\times$ and black $+$ data points were measured for $18$ by $18$ unit cells for the red and black trajectory through the Brillouin zone depicted in (b). In contrast to a usual honeycomb reciprocal lattice vector structure, the gauge for the circuit Brillouin zone is chosen to be quadratic (straight white), which upon folding takes the form of a distorted hexagon (dashed white). The heat map of admittance in reciprocal space stresses the dominant low spectral regime around the $K/K'$ points, hence dominating the impedance read-out.} 
		\label{fig:graph}
	\end{figure*}
	
	{\it Admittance band analysis.} We label each node in the circuit by an index $j$, where the voltage at that node, $V_j$, is measured with respect to ground. The input current, which defines the current flowing into the circuit at that node from the outside world, is denoted by $I_j$. With this, we are able to arrange the components $I_j$ and $V_j$ in a vector form linked by using Ohm's and Kirchhoff's law (Appendix A), 
	\begin{align}\label{eq:grounded_circuit_laplacian}
	\nvec{I} = J(\omega) \, \nvec{V}.
	\end{align}
	$J$ denotes the grounded circuit Laplacian and $\omega$ the AC driving frequency of the excitation current applied to the circuit~\cite{ustopo}, which takes the role of an external parameter.
	The response of the system to a given input current signal is governed by the eigenstates of $J$. The impedance resonance frequencies $\omega_{\text{res}}^{(n)}$ are the roots of the Laplacian's eigenvalues $j_n(\omega)$, $n \in \{1,\dots, \text{dim}[J]\}$~\cite{ustopo}.
	The circuits we investigate are composed of a repeating minimal set of $M$ nodes and conductances, which together we call the circuit unit cell. The nodes of a circuit representing a $D$-dimensional network can be labeled by two indices $j \equiv (\rho,\alpha)$, where $\rho$ is an index denoting the unit cell and $\alpha \in \{1,\dots, M \}$ the nodes within a unit cell. $D$ specifies the synthetic dimension of circuit lattice periodicity, which is determined by the maximum number of linearly independent Bravais lattice vectors $\nvec{R}_\rho$.
	Note that, as one central difference to a solid state lattice, there is no fixed length or orientation of the Bravais vectors, as the circuit lattice truly is a graph, and as such solely determined by lattice connectivity. This implies an equivalence class of different choices of Bravais vectors, and thus gives an additional gauge for the circuit lattice network (Appendix~B). 
	Once we fix a Bravais vector gauge $\{\nvec{R}_{\rho}\}$, we can diagonalize a translationally invariant $J$ by performing a Fourier transform to $D$-dimensional reciprocal space $\nvec{k}$ into $M$-dimensional block matrices
	\begin{align}\label{eq:fourier_transform_of_periodic_laplacian}
	J_{\alpha\beta}(\nvec{k},\omega) = \sum_\rho J_{\alpha\beta}( R_{\rho},\omega) \,  \exp\left[-i \, \nvec{R}^{\top}_\rho \, \nvec{k} \right].
	\end{align}
	To find the eigensystem of the Laplacian matrix, and hence the admittance band structure, we diagonalize the block matrices $J_{\alpha\beta}(\nvec{k},\omega)$. 
	The Laplacian matrix in reciprocal $\nvec{k}$-space forms an irreducible representation of the translation group. The admittance band structure can then be seen as the irreducible representation of the space group incorporating the periodic circuit configuration in graph space. \\

	
	\begin{figure*}
		\includegraphics[width=\linewidth]{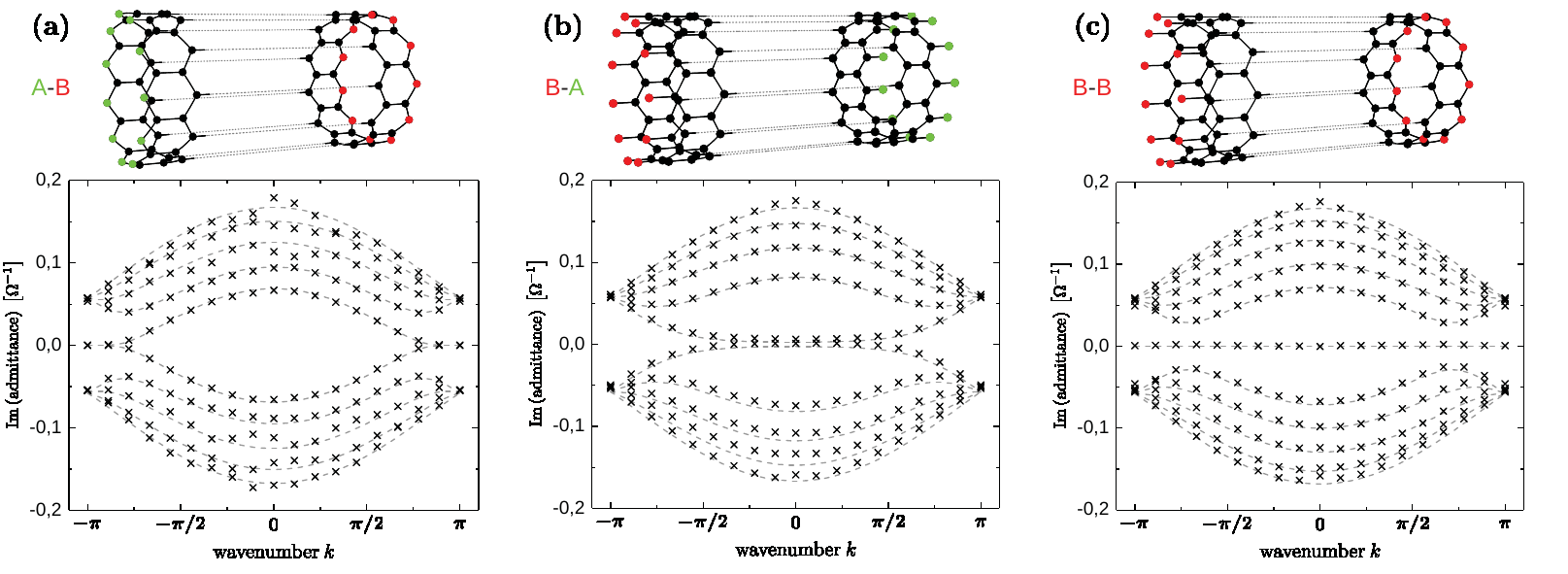}
		\caption{Admittance band analysis of different graphene circuit edge terminations given by (a) A-B zigzag, (b) B-A bearded, and (c) B-B bearded/zigzag termination, where A (green) and B (red) used in the schematic honeycomb geometry label the different nodes of the circuit unit cell. The admittance band structure data (black $\times$ points) derive from an cylindric circuit geometry with 18 unit cells along the periodic and 5 along the open direction, where the grey dashed lines highlight the theoretical expectation for the bands. Depending on the termination, flattened spectral admittance features are visible whose eigenstates localize at the termination. For the Brillouin zone gauge chosen in Fig.~\ref{fig:graph}, (a) yields a flat spectrum at zero admittance for $\vert k \vert > 2\pi/3$ while (b) shows the complementary flattening for $\vert k \vert < 2\pi/3$. Independent of the Brillouin zone gauge, the B-B terminated circuit in (c) exhibits a flat admittance band.}
		\label{fig:edge}
	\end{figure*}

	{\it Admittance band measurement.} 
	We apply an input current at one specified node of the circuit and measure the response of the circuit given by the complete voltage vector with respect to that input current. If we apply an input current at node $j$, we can compute the impedances
	\begin{align}\label{eq:calculation_of_G}
	G_{i j} = V_{i}^{(j)}/{I_j} = J_{i j}^{-1}.
	\end{align}
	$V_{i}^{(j)}$ represents the voltage measured at node $i$ when the only input current to the circuit is given by $I_j$. 
	As the matrix $G$ is the inverse of the circuit Laplacian $J$~\cite{ustopo}, the complex valued admittance eigenvalues are obtained by inverting the eigenvalues of $G$. 
	Note that by analogy, such site-resolved measurements are out of reach in generic transport or scattering experiments on physical crystals.
	For a randomized system of $N$ nodes, the measurement procedure of exciting one node and measuring the whole voltage profile needs to be repeated $N$ times to recreate the matrix $G$ by use of~\eqref{eq:calculation_of_G}, where each of the $N$ measurement processes features an input current at a different node. If we are dealing with a fully periodic system, however, only $M$ nodes are inequivalent. In this case, we thus  restrict ourselves to repeating the outlined measurement procedure $M$ times, where each sublattice needs to be supported once (Appendix C). The data of the voltage and the current vector is then Fourier transformed to reciprocal space, and the $(M \times M)$ impedance matrix is recovered for each $\nvec{k}$ by use of~\eqref{eq:calculation_of_G}. We determine the complex Laplacian matrix and its eigenvalues for each $\nvec{k}$ separately, and thus restore the band structure. The measurement principle readily extends to the case of open boundary conditions for any synthetic circuit dimension. 
	
	{\it Honeycomb circuit.} As introduced in Ref.~\onlinecite{ustopo}, with a unit cell depicted in the inset of Fig.~\ref{fig:graph}(a), we consider the analogue to a honeycomb structure in a circuit network.
	We thus have $M=2$ and three equivalent capacitive conductances $C$ per node to other nodes: 
	\begin{align}\label{eq:laplacian_graphene_k_space}
	J_{\text{hc}}(\nvec{k}) = & \,  \im \omega \, \big[ \left(3 \, C -\frac{1}{\omega^2 \, L}\right) \, \mathbbm{1}\nonumber\\ &- C \, \left(1+ \cos(k_x) + \cos(k_y) \right) \, \sigma_x\nonumber\\ &- C \left(\sin(k_x) + \sin(k_y)\right) \, \sigma_y \, \big],
	\end{align}
	yielding a two-band structure given by 
	\begin{align}\label{eq:eigenspectrum_graphene}
	&j_{\text{hc}}^{(\pm)}(\nvec{k}) = \im \omega \bigg[\left(3 \, C -\frac{1}{\omega^2 \, L}\right)  \nonumber \\
	&\pm C \, \sqrt{3+2 \cos(k_x) + 2 \cos(k_x-k_y) +2 \cos(k_y)} \bigg].
	\end{align}
	AC-driving with the characteristic resonance frequency $\omega_0 = 1/\sqrt{3 \,L \, C}$ eliminates the offset proportional to identity and symmetrizes the honeycomb lattice spectrum around zero admittance.
	In the absence of disspative losses such as imposed by serial resistances, the spectrum is purely imaginary.

	For the experimental implementation, we devise standard printed circuit boards (PCB), and fit them with commercially available electronic components (Appendix D). The PCB modules for the honeycomb circuit are designed to contain 6 by 6 unit cells with the option of selecting specific components at the edge termination. We serially connect the edges in both spatial dimensions to fuse several PCB modules and set the circuit termination to either provide periodic or open boundary conditions. The driver current is fed into a particular sublattice site from ground.
	The measurements of the AC voltages are done by Stanford Research 530 Lock-In Amplifiers. The driving current is detected as a voltage drop through a shunt resistor. The driving frequency is set to the respective operational resonance frequency, which is identified in the impedance spectrum recorded by a BK Precision 894 LCR-meter. The reconstructed band structure measurement is summarized in Fig.~\ref{fig:graph}. As seen, the data is in good correspondence to the theoretical prediction~\eqref{eq:eigenspectrum_graphene}. The deviations of large admittance eigenvalues from theory are greater due to reduced excitation of the corresponding eigenstates (Appendix D).
	The red/black data points in Fig.~\ref{fig:graph}(a) correspond to the red/black path taken in the Brillouin zone as shown in Fig.~\ref{fig:graph}(b). To illustrate the Bravais gauge, we have picked the Brillouin zone to take the form of a square (brick wall type) which, upon suitable reciprocal folding, appears like a distorted hexagon (Appendix C). 
	While the spectrum from~\eqref{eq:eigenspectrum_graphene} is gauge invariant, the map onto wave vector momenta is not, leading to the distorted spectrum for the chosen gauge.

	{\it Open boundary termination.} We adjust the honeycomb circuit PCBs to exhibit open boundary conditions in one brick wall direction while keeping periodic boundary conditions for the other. Due to two sublattice components and two choices of termination of the resulting cylindric geometry, different settings can be investigated. Fig.~\ref{fig:edge} shows the predicted and measured admittance band structure for different choices of termination, where we put an emphasis on those exhibiting flat surface admittance modes. Viewed together, Fig.~\ref{fig:edge}(a) and Fig.~\ref{fig:edge}(b) display one complete flat band of admittance eigenvalues, which is doubly degenerate because of the two identical edges. The flat band splits into a regime $\vert k \vert > 2\pi/3$ and $\vert k \vert < 2\pi/3$ between the A-B zigzag and bearded termination, respectively. For the B-B bearded/zigzag termination, Fig.~\ref{fig:edge}(c) displays a non-degenerate flat band where, if it were resolved with respect to the two edges, the same distribution between the bearded and zigzag edge would be observed as for Fig.~\ref{fig:edge}(a) and Fig.~\ref{fig:edge}(b).
	
	\begin{figure}
		\includegraphics[width=\linewidth]{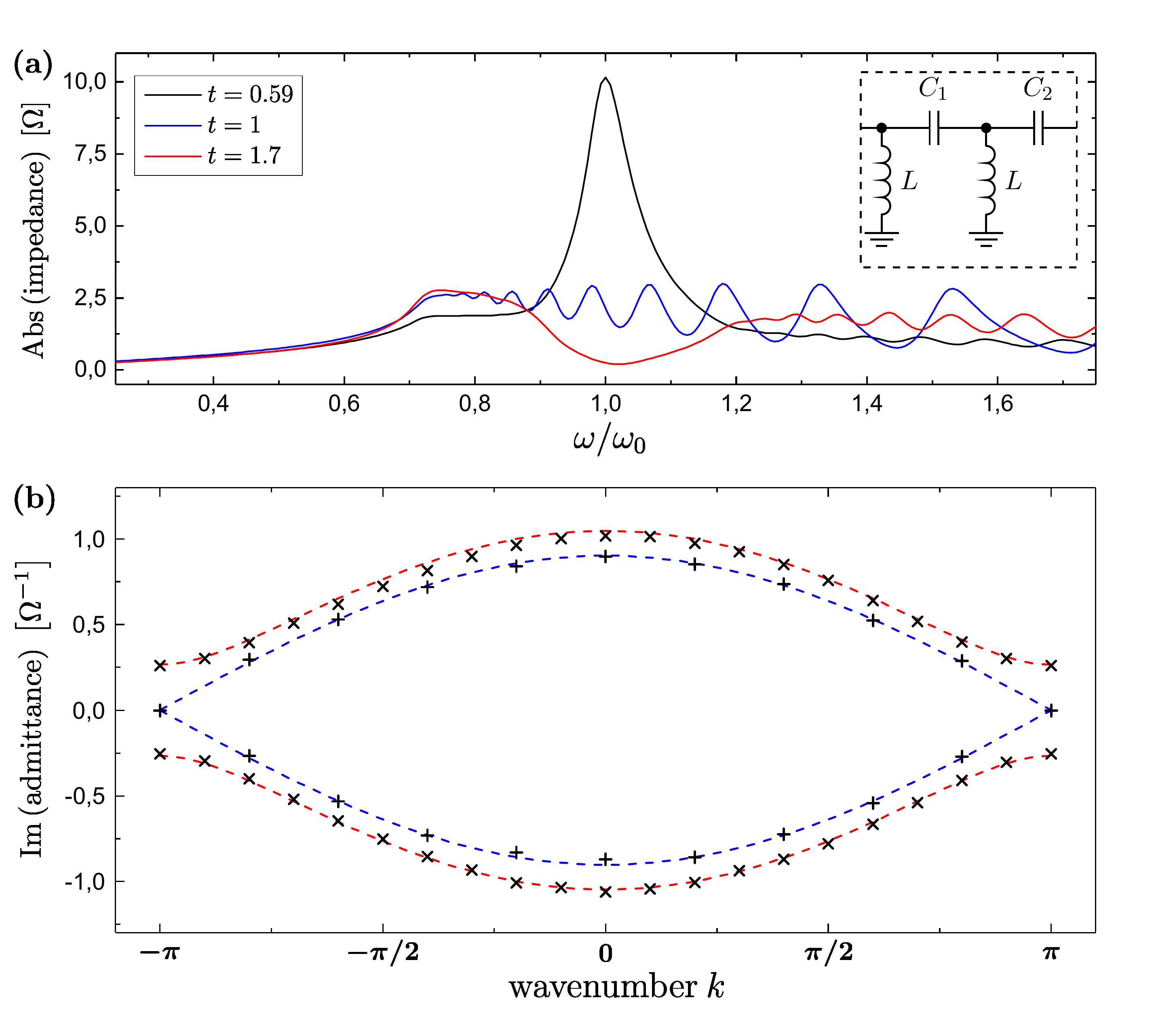}
		\caption{(a) Absolute value of impedance as a function of AC frequency $\omega$ for the open boundary SSH circuit (unit cell depicted at the upper right inset) at the critical value $t=1$ (blue, 10 unit cells), $t=1.7$ (red, 19 unit cells) contained in the topologically trivial regime $t>1$, and $t=0.59$ (black, 20 unit cells) contained in the non-trivial regime $t<1$. At $\omega=\omega_0$, the topolectrical boundary resonance (TBR) related to the topological SSH midgap states is resolved. (b) Admittance band structure measured for the periodic SSH circuit. For $t=1$ ($+$ sign, continuum theory curve in dashed blue), the band structure appears critical at $k=\pi$. In the periodic case, $t=1.7$ and $t=0.59$ yield the same bands ($\times$ sign, continuum theory curve in dashed red) due to spectral self-similarity under $t\rightarrow 1/t$, with an admittance gap at $k=\pi$.} 
		\label{fig:ssh}
	\end{figure}   
	
	{\it Topological phase transition.} The admittance band measurement we propose also allows to track the bulk topological phase transition of a topolectrical circuit. As its most elementary representative, we study the Su-Schrieffer-Heeger (SSH) circuit, an $M=2$ one-dimensionally connected circuit whose admittance band structure corresponds to that of the SSH tight binding model for polyacetylene~\cite{ustopo,PhysRevLett.25.919,su-79prl1698}. The conductances are given by capacitors with capacitance $C_1$ inside the unit cell (intracell) and $C_2$ between adjacent unit cells (intercell) described by the paramter $t=C_1/C_2$. Each node is also connected to ground by an inductor with inductance $L$ (see inset in Fig.~\ref{fig:ssh}(a)). The SSH PCBs are designed to contain ten unit cells, with the option to have different edge terminations and to stack several circuit boards by connecting them in series. The circuit Laplacian is given by~\cite{ustopo}
	\begin{align}\label{eq:laplacian_SSH_k_space}
	J_{\text{SSH}}(k) & = \ \im \omega \, \big[ \left(C_1+C_2 -\frac{1}{\omega^2 \, L}\right) \, \mathbbm{1} \nonumber\\
	- & \, \left(C_1 +C_2 \, \cos(k)\right) \ \sigma_x - \left(C_2 \, \sin(k)\right) \ \sigma_y \, \big],
	\end{align}
	yielding the admittance band structure
	\begin{align}\label{eq:eigenspectrum_SSH}
	j_{\text{SSH}}^{(\pm)}(k) = \im \omega \bigg[  &\left(C_1+C_2 -\frac{1}{\omega^2 \, L}\right)  \nonumber \\
	&\pm \sqrt{C_1^2+C_2^2+2\,C_1 \, C_2 \, \cos(k)} \bigg].
	\end{align}
	Fig.~\ref{fig:ssh}(a) depicts an open boundary impedance measurement for the parameters $t=1$ as well as $t=0.59$ and its inverse $t=1.7$. In the topologically non-trivial regime $t<1$, at $\omega_0=1/\sqrt{L(C_1+C_2)}$, the circuit exhibits an admittance midgap state at the boundary, which manifests as an impedance peak. In the dissipationless limit and for an exact zero admittance SSH midgap state, this peak would be divergent, but in reality becomes damped due to serial circuit resistance and component disorder~\cite{ustopo}. This peak is absent for $t>1$. Fig.~\ref{fig:ssh}(b) shows the reconstructed bulk admittance band structures. Because of the duality under $t\rightarrow 1/t$, the bulk spectrum is identical for $t=0.59$ and its inverse, showing a bulk admittance gap. The phase transition occurs at $t=1$, where the admittance gap closes at $k=\pi$. 

	{\it Conclusions and outlook.} Electric circuit networks, together with the admittance band measurement protocol developed in our work, establish a promising platform for the design, engineering, and measurement of tight-binding models. In comparison to alternative frameworks of synthetic matter, electric circuits offer unique advantages. First, electric circuits are placed in the infinite tight-binding limit, and as such arbitrarily scalable. Second, the circuit boundary conditions can be conveniently switched between open and periodic, allowing to investigate bulk band properties and edge states in the same experimental sample. Third, while we have not yet exploited it in this work, arbitrary longer ranged hopping can be straightforwardly considered, alongside with realizing lattices of arbitrary dimension and connectivity. Here, the graph property of electric circuits will allow for the implementation of symmetries independent of the physical embedding space which are in part inaccessible to physical crystals. Together with their unprecedented feasibility and accessibility, electric circuit networks promise to yield fundamental insights into topological band structures~\cite{ningyuan2015time,PhysRevLett.114.173902,ustopo,ustopo2,kaifa,serra} and beyond. 
	
	
	{\it Note added.} Upon completion of this manuscript, we became aware of a contemporaneous work~\cite{schustersimon} providing an experimental realization for a Weyl circuit~\cite{ustopo}. An inductive nodal measurement is performed to reconstruct energy band dispersion which is not rigid, i.e. sensitive to the energy offset, while we reconstruct rigid admittance bands, i.e. insensitive to the grounding adjustment of admittance. 
	
	\section{Acknowledgments}
	\begin{acknowledgments}
		R. Thomale thanks S.~Huber, T.~Neupert, M.~Rechtsman, A.~Szameit, and V.~Vitelli for discussions. The work in W\"urzburg is supported by the European Research Council (ERC) through ERC-StG-TOPOLECTRICS-336012 and by the German Research Foundation (DFG) through DFG-SFB 1170. 
	\end{acknowledgments}

	


\appendix

	\section{Appendix A: Definition of the Circuit Laplacian}
\label{app:lap}
Building up on the method of nodal analysis, let us consider an electrical network consisting of $N$ nodes, which are linked by linear elements in the form of serial resistors $R$, inductors $L$ and capacitors $C$. The nodes are labeled by the index $j = 1, \dots , N$ while the ground always retains the index 0. We measure the voltage $V_j$ with respect to the ground node $V_0 = \SI{0}{\volt}$ and allow a current $I_j$ to enter the system at the $j$th node. Two individual nodes of the circuit, $j$ and $l$ can be connected by an admittance $g_{j l}$, which is zero if they are not connected directly. $I_{j l}$ with two indices shall denote the current running from node $j$ to a directly connected node $l$. By Kirchhoff's rules the input current at the $j$th node is then given by
	\begin{align}\label{eq:derivation_Laplacian}
	I_j = \sum_{l=0}^N I_{j l}  ,
	\end{align}
	where $I_{j l} = g_{j l} \, (V_j - V_l)$. Inserting this into above formula yields the circuit Laplacian
	\begin{align}
	I_j &= \sum_{l=0}^N g_{j l} \, (V_j - V_l)  \nonumber \\ 
	&= \sum_{l=1}^N \left( \sum_{m=1}^N g_{j m} \, \delta_{j l} + g_{j 0} \, \delta_{j l} - g_{j l} \right)  V_l \nonumber \\
	&=:\sum_{l=1}^N J_{j l} V_l .
	\end{align}
	We call $J$ the grounded circuit Laplacian with the matrix respresentation $J_{j l}$, as its continuum analogue is the Laplacian operator~\cite{ustopo}. We relabel the lattice sites of the circuit in agreement with the main text, $j=(\rho, \alpha)$, $l=(\sigma, \beta)$, to be able to investigate periodicity in the circuit. Translational invariance causes the Laplacian to only depend on the difference between Bravais lattice vectors $R_{\rho}$ and $R_{\sigma}$,
	\begin{align}
	J_{\rho,\alpha;\sigma,\beta}(\omega) = J_{\alpha \beta}(R_{\rho} - R_{\sigma}, \omega),
	\end{align} 
	and hence allows for a description in reciprocal space $J_{\alpha \beta}(\nvec{k})$. The Fourier transformation defined in equation \eqref{eq:fourier_transform_of_periodic_laplacian} provides the mapping of the Laplacian spectrum to the wave vector $\nvec{k}$, which ultimately diagonalizes the Laplacian in reciprocal space and hence establishes the admittance band structure of an electric circuit array. 
	
	In analogy to tight-binding Hamiltonians, modifying the connections among the nodes corresponds to a change of the hopping elements. Note that in the Laplacian formalism, this does also affect the diagonal terms of the total node conductance~\cite{ustopo}. Moreover, connections from each node to ground are represented by a diagonal matrix. In tight-binding language, those terms are taken into account as on-site potentials.
	
	\section{Appendix B: Gauge Symmetry of reciprocal space}\label{app:gau}
\begin{figure*}
		\includegraphics[width=0.9\linewidth]{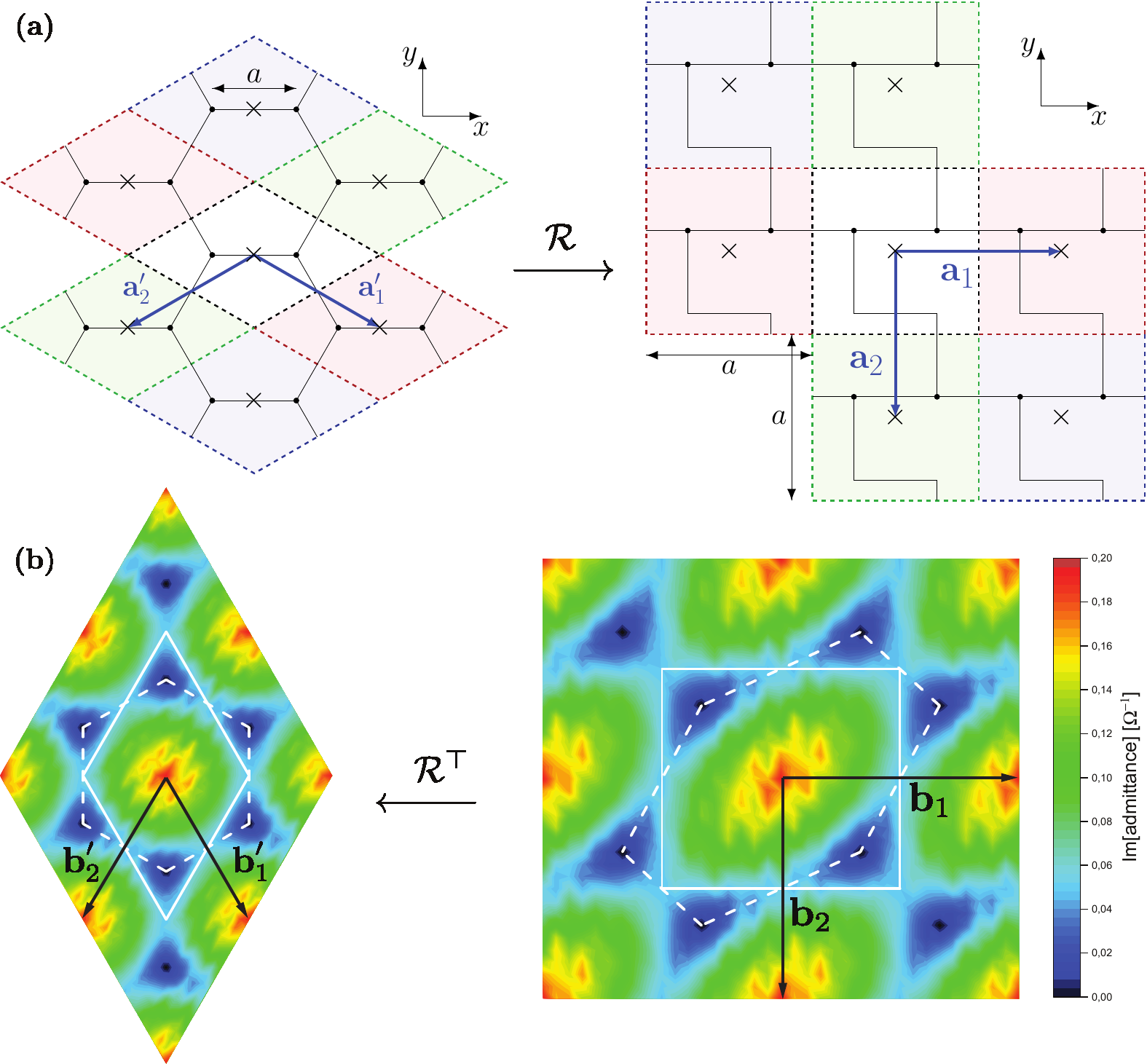}
		\caption{(a) Illustration of the circuit gauge transformation from the honeycomb setting to the brick wall type configuration in graph space. The connectivity between adjacent nodes remains invariant, while the shape of the unit cell is distorted. The transformation matrix $\mathcal{R}$ transforms the primitive lattice vectors from $\nvec{a}_1', \nvec{a}_2'$ to $\nvec{a}_1, \nvec{a}_2$. The different colouring of the unit cells are a guide to the eye. (b) Transformation of the measured band structure from the quadratic to the hexagonal Brillouin zone configuration along with the primitive reciprocal vectors from $\nvec{b}_1, \nvec{b}_2$ to $\nvec{b}_1', \nvec{b}_2'$. The spectrum of the Laplacian remains invariant, while solely the shape of the representation in reciprocal space is altered by the transformation.}\label{fig:honeycomb_gauge}
	\end{figure*}

	As the electric circuit is a graph, there cannot be any associated length scales with the network lattice structure. As alluded to in the main text, we acquire an additional gauge symmetry of choosing the Bravais lattice vectors $\{\nvec{R}_\rho\}$. We want to focus on demonstrating this and further derive implications on reciprocal space, by exemplifying the gauge choice for the honeycomb circuit lattice. Consider the conventional choice of primitive vectors for the honeycomb lattice, $\nvec{a}'_1 = a (3, -\sqrt{3})/2$ and $\nvec{a}'_2 = a (-3,-\sqrt{3})/2$ (see fig. \ref{fig:honeycomb_gauge}(a), left), where $a$ is the lattice constant, which is set to $1$. The corresponding reciprocal lattice vectors therefore are $\nvec{b}_1'=2\pi(1,-\sqrt{3})/3$ and $\nvec{b}_2'=2\pi(-1,-\sqrt{3})/3$. As stressed before, we select a different gauge of primitive vectors, $\nvec{a}_1 = (1,0)$ and $\nvec{a}_2 = (0,-1)$ (Fig.~\ref{fig:honeycomb_gauge}(a)), with their reciprocal complements $\nvec{b}_1 = 2\pi (1,0)$ and $\nvec{b}_2 = 2\pi (0,-1)$. Both choices of lattice vectors are connected by the transformation
	\begin{align}
	\mathcal{R}: \ \nvec{a}_d' \rightarrow \nvec{a}_d = \mathcal{R} \, \nvec{a}_d'&, \qquad d=1,2 \nonumber\\
	\text{with} \ \ \mathcal{R} = \begin{pmatrix} 1/3 & -1/\sqrt{3}\\ 1/3 & 1/\sqrt{3} \end{pmatrix}&.
	\end{align}
	This converts the honeycomb lattice into the brick wall-type configuration. As the connections are unchanged (Fig.~\ref{fig:honeycomb_gauge}), the circuit's observable behaviour is invariant, and $\mathcal{R}$ is a gauge transformation.
	
	From the definition of the reciprocal lattice vectors, $\nvec{a}_d^\top \nvec{b}_{d'} = 2\pi \, \delta_{d,d'}$ it follows that the transformation $\mathcal{R}$ acts on them as
	\begin{align}
	\nvec{b}_d' \rightarrow \nvec{b}_d = (\mathcal{R}^{-1})^\top \, \nvec{b}_d' .
	\end{align}
	Therefore, we recover the familiar shape of the honeycomb model band structure using the inverse transformation $\mathcal{R}^\top$ in reciprocal space, i.e. $\nvec{k} \rightarrow \mathcal{R}^\top \, \nvec{k}$. The action of the transformation on the experimental data is shown in Fig.~\ref{fig:honeycomb_gauge}(b), where the hexagonal structure of the Brillouin zone is recovered. 
	
	\section{Appendix C: Detailed remarks on the circuit measurement}\label{app:mea}
	The procedure in the main text to obtain the admittance band structure involves a current fed from ground to one single node, and the subsequent measurement of its voltage response at the individual circuit nodes. Alternatively, it is also possible to apply input currents to several nodes in the circuit during one measurement process, e.g. by attaching two nodes (instead of one node and ground) to a current or voltage source. Let $m$ denote the index of independent measurements. Recast as a system of linear equations, one then needs to determine the solution of a system of $N^2$ equations given by
	\begin{align}
	\nvec{V}^{(m)} = G^{(m)} \, \nvec{I}^{(m)}, \qquad m=1, \dots,N.
	\end{align}
Thereby one needs to apply the currents such that the resulting equations are linearly independent. If working with voltage or current sources which are only connected to the circuit board and {\it not} to ground, one needs to be aware of the additional constraint $	\sum_j V_j = 0$.
It implies, that the $\nvec{k}=0$ component of the voltage and the current vector is fixed to zero. In this configuration, we are therefore not able to excite the eigenmode corresponding to $\nvec{k}=0$ and cannot recover the corresponding $\nvec{k}=0$ eigenvalue. 

	In addition to the real space placement of the driving contacts, we further specify the role of the frequency parameter in the context of the Laplacian formalism. The circuit lattice network exhibits resonance frequencies as the roots of the eigenvalues of the Laplacian,
	\begin{align}
	j_n(\omega = \omega_\text{res})= 0.
	\end{align}
	When the frequency parameter is set to such a resonance frequency, the admittance of at least one eigenmode is zero, and the impedance of the circuit diverges while the input current drops to zero. Furthermore, the voltage vector is an eigenvector $\psi_n$ of the Laplacian matrix $J$. In the experimental circuit setup, the circuit will always feature parasitic effects such as serial resistances. Due to that, by setting the AC driving frequency $\omega$ to a resonance, we encounter smoothened impedance peaks with finite height, which results from eigenvalues acquiring a small deviation from zero. The voltage vector will consequently be a superposition of the available eigenmodes, which then needs to be analyzed. Using the circuit Laplacian, we can determine the response of the system to a given current input vector $\nvec{I}$. The voltage vector resulting from this excitation can be expanded in terms of the eigenvectors of the Laplacian,
	\begin{align}
	\nvec{V} = \sum_{n = 1}^N u_n \, \psi_n
	\end{align}
	with the complex coefficients
	\begin{align}
	u_n = \frac{1}{j_n(\omega)} \, \psi_n^{\top} \nvec{I} .
	\end{align}
	They provide the degree to which an eigenmode is excited at a given external AC frequency $\omega$. This excitation weight thus is proportional to the inverse of the corresponding eigenvalue as well as to the inner product of the eigenvector and the current vector. 

This finding is useful for interpreting the experimental implementation of the measurement. As eigenstates attached to admittances with greater absolute value $\left\vert j_n(\omega)\right\vert$ are less excited during the measurement, the signal-to-noise ratio for those values is reduced, leading to larger statistical deviations of the corresponding data points from theory. By tuning the frequency, we can alter the eigenvalues and adjust the excitation of specific eigenmodes. Experimentally, it is therefore convenient to choose a frequency which ensures that the excitation of eigenmodes of the upper band is the same as of the lower band, such that no distortion between the bands is generated and the mean excitation of eigenmodes is maximized.
		For the SSH and the graphene circuit this means that we choose their resonance frequency $\omega_0$. The bands are then symmetric around zero admittance. Any persistent asymmetry in our circuits band structure arises as an artefact of asymmetric excitation due to parasitic effects, such as additional dissipative components. 

	\begin{figure}[t]
		\includegraphics[width=\linewidth]{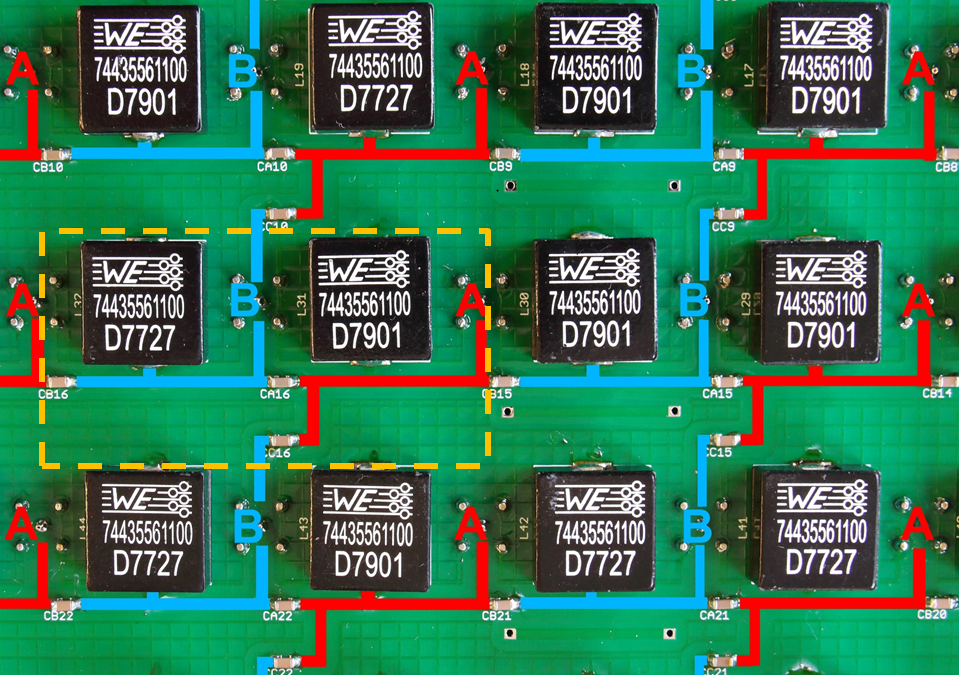}
		\caption{Printed circuit board for the graphene circuit. Labels highlight the circuit lattice connectivities imposed on the individual circuit elements. The yellow dashed square denotes a single unit cell in correspondence to the inset in Fig.~\ref{fig:graph}.} 
		\label{fig:board}
	\end{figure}   
	
	\section{Appendix D: Experimental implementation}\label{app:exp}
	The circuits were designed to resonate close to the maximum frequencies of our lock-in amplifiers of 100 kHz to enhance signal-to-noise ratios. The design strategy is then summarized as follows: The unavoidable serial resistance of the inductors should be kept as small as possible to obtain spectrally sharp features. This requires the values of the inductances also to be kept small as the two quantities scale together. As the resonance frequency $\omega_0$ is inversely proportional to both $L$ and $C$, this implies that the capacitances should be chosen as large as possible under practical considerations such as commercial availability and PCB compatible design form.
	
	To preserve translational symmetry the scatter of the absolute values of the circuit elements needs to be smaller than typical tolerances of commercially available components ensure. To this end all components were pre-characterized by a BK Precision 894 LCR-meter. Finally, the PCBs need to be designed with sufficient line spacing and a magnetic shielding to suppress spurious inductive coupling.
	
	The above considerations resulted in the following choice of components: All boards were fit with SMD Flat Wire High Current Inductors 74435561100, nominal values $L=\SI{10}{\micro\henry}$ and $R_\text{DC}=\SI{6.9}{\milli\ohm}$. For the SSH circuits we further used Murata Multilayer Ceramic Capacitors (GCG31CR71E475JA01L and C1210X825K3RATU). For the $t=0.59/1.7$ SSH board the components were selected to absolute values of $L = \SIrange{10.58}{10.60}{\micro\henry}$, $C_1 = \SIrange{4.32}{4.34}{\micro\farad}$ and $C_2 = \SIrange{7.28}{7.32}{\micro\farad}$, which sets the resoncance frequency to about $\SI{15}{\kilo\hertz}$. For the $t=1$ SSH board the components were taken to be  $L = \SIrange{10.62}{10.64}{\micro\henry}$, $C_1 = C_2 = \SIrange{4.32}{4.34}{\micro\farad}$. For the graphene-type PCBs (Fig.~\ref{fig:board}), we employed Kemet Multilayer Ceramic Capacitors (C1206C104F3GACTU), the components sorted to $L = \SIrange{10.00}{10.20}{\micro\henry}$ and  $C = \SI{0.1}{\micro\farad{} \pm{} 1\percent}$. The latter sets the resonance frequency to about $\SI{90}{\kilo\hertz}$.


\end{document}